\documentclass[12pt]{article}
\usepackage{amssymb}
\usepackage{epsfig}
\newcommand{\beqn}{\begin{eqnarray}}
\newcommand{\eeqn}{\end{eqnarray}}
\newcommand{\eq}[1]{(\ref{#1})}
\newcommand{\tr}{\mathop{\rm Tr}}
\newcommand{\NP}[3]{{\em Nucl. Phys. }{\bf #1}, #3 (#2)}
\newcommand{\NPPS}[3]{{\em Nucl. Phys. Proc. Suppl. }{\bf #1}, #3 (#2)}
\newcommand{\PL}[3]{{\em Phys. Lett. }{\bf #1}, #3 (#2)}
\newcommand{\PRL}[3]{{\em Phys. Rev. Lett. }{\bf #1}, #3 (#2)}
\newcommand{\PRep}[3]{{\em Phys. Rep. }{\bf #1}, #3 (#2)}
\newcommand{\PR}[3]{{\em Phys. Rev. }{\bf #1}, #3 (#2)}

\newcommand{\abstracts}[1]{{
\centering{\begin{minipage}{13.0truecm}
\normalsize\baselineskip=15pt \centerline{\footnotesize
ABSTRACT}\vspace*{0.3cm}
\parindent=20pt #1
\end{minipage}}\par}}

\begin{document}

\begin{center}
{\baselineskip=24pt {\Large \bf Anatomy of the lattice magnetic
monopoles}\\

\vspace{1cm}

{\large
V.G. Bornyakov$^{\rm a,b}$, M.N. Chernodub$^{\rm c}$,
F.V.~Gubarev$^{\rm c,d}$, \\
M.I. Polikarpov$^{\rm c}$, T. Suzuki$^{\rm e}$, A.I. Veselov$^{\rm c}$,
V.I.~Zakharov$^{\rm d}$}
}

\vspace{.5cm}
{\baselineskip=16pt
{ \it

$^{\rm a}$ Deutsches Elektronen-Synchrotron DESY
    \& NIC, D-15735 Zeuthen, Germany\\
$^{\rm b}$ Institute for High Energy Physics, Protvino 142284,
    Russia\\
$^{\rm c}$ Institute of Theoretical and  Experimental Physics,
    Moscow, 117259, Russia\\
$^{\rm d}$ Max-Planck Institut f\"ur Physik, F\"ohringer Ring 6,
    80805 M\"unchen, Germany \\
$^{\rm e}$ Institute for Theoretical Physics, Kanazawa University,
    Kanazawa 920-1192, Japan
}
}
\end{center}

\vspace{5mm}

\abstracts{We study the Abelian and non-Abelian action density
near the monopole in the maximal Abelian gauge of $SU(2)$ lattice
gauge theory. We find that the non-Abelian action density near the
monopoles belonging to the percolating cluster decreases when we approach
the monopole center. Our estimate of the monopole radius is
$R^{mon} \approx 0.04 \,\, fm$.}

%=============================================================================
\section{Introduction}

Confinement of color in QCD implies that the color field of
external quarks is squeezed into a tube connecting the quarks
(provided that the distance between the quarks is large enough).
Similarly, the ordinary magnetic field cannot penetrate
superconductors and the dual superconductor model of confinement
\cite{confinement} makes this analogy manifest. The model assumes
condensation of magnetic monopoles in QCD, similar to the
condensation of charged Cooper pairs in superconductor. The
monopole confinement mechanism is confirmed in $SU(2)$ lattice
gauge theory  by many numerical calculations, for a recent review
see, e.g., \cite{reviews}. Microscopically, the condensation of
the monopoles can be understood as percolation of a monopole
cluster. And, indeed, it was observed  that there exists always a
big percolating cluster which is responsible for confinement
\cite{percolation}. Since the percolating cluster may seemingly
have any size we will call it infrared (IR). On the other hand,
there are also many small, or ultraviolet (UV) clusters which are
usually viewed as lattice artifacts \cite{percolation}.

Although there is a lot of data on the lattice monopole the
understanding of the monopole dynamics in terms of the continuum
theory is far from being complete at the moment. Qualitatively,
there are two ways of looking at the monopoles in non-Abelian
theories. First, one can think in terms of an analogy with the 't
Hooft-Polyakov monopoles \cite{thooftpolyakov} which are classical
solutions to the Yang-Mills equations with a triplet of matter
fields. However, there are no matter fields in QCD. As a result,
one rather changes the strategy of defining the monopoles
\cite{thooftt}.  Namely, they can be defined  as purely
topological defects, with no direct relation to the density of the
non-Abelian action. According to the original idea of
Ref.~\cite{thooftt} in case of $SU(2)$ gauge group one can choose
any vector in color space and (partially) fix the gauge by
rotating the vector to the third direction.  Such  gauge fixing
fails when all the components of the vector vanish at some point.
The crucial observation is that vanishing of a vector gives three
conditions which in the D=4 case define line-like defects, that is
the monopole trajectories. The success of the monopole confinement
model depends in fact on the particular choice of the gauge. The
observation might imply that a purely topological definition,
devoid of any dynamic content is not in fact adequate. The so
called maximal Abelian gauge and the corresponding projection
\cite{reviews} turns out to be the most carefully studied and very
successful.  Since the Abelian projection emphasizes the role of
the Abelian-like field configurations this might be an indication
that at large distances the lattice monopoles are similar to the
Abelian or Dirac monopoles.

To get insight into the dynamics of the lattice mo\-no\-po\-les we
will concentrate here on measuring the full non-Abelian and
Abelian actions at the centers of the monopoles. Actually, this
kind of measurements have been reported earlier \cite{monphys}.
Namely, it was shown that the non-Abelian action on the plaquettes
close to the monopole trajectory is larger than the average
plaquette action, $S$. It is easy to realize, however, that if
this were true at arbitrary small distances the monopoles would be
strongly suppressed by the action factor and could not condense,
see, e.g., \cite{shiba2}. In this note we report on the
measurements which demonstrate for the first time that the above
mentioned excess of the action  goes down for smaller lattice
spacing, or larger $\beta$. A crucial novel point  is that we
distinguish between the monopoles belonging to the UV and IR
clusters and the statement on the decreasing of the action refers
to the IR monopoles only. In this sense, the structure of the IR
and UV monopoles turns out to be different and one can say that
the monopoles in the IR clusters are condensed due to their
special anatomy.

The separation of the monopole ensemble on IR and UV clusters is
unambiguous for large enough lattices. The distribution of the
cluster lengths clearly shows~\cite{percolation} that each
monopole configuration contains typically one large IR cluster and
a lot of small UV clusters separated by clearly observed gap. Thus
we do not need to introduce any artificial mass scale to
distinguish between IR and UV clusters.

The most important question to be addressed here is the estimation of
the size of the monopoles.
%We define the size as the distance where the
%action density has maximal slope (see Sect. 3).
Our definition of the monopole size will be described in Sect. 3
As we shall see, the
size of the monopole turns out to be rather small numerically. This
observation supports speculations on the existence of a numerically
large mass scale in the non-perturbative physics, see, e.g.,
\cite{sumrules,scales} and references therein. On the other hand
our results show that the size of the Abelian monopoles is much smaller
than the distance between the monopoles. Thus the monopole cores are not
overlapping and the system can be tractable as a dilute gas.
%
%This feature distinguishes the dual superconductor model of the QCD vacuum
%from, $e.g.$, the instanton models in which the sizes of the instantons
%and the distances between them are of the same order.

In the next section we will summarize the current views on the
anatomy of the monopoles. In section 3 we present our data and
discuss their implications. It occurs that due to the finite size
the monopoles in gluodynamics are condensed at any value of the
bare coupling. In compact QED, on the other hand, where monopoles
are point-like, the critical coupling, separating confinement
and deconfinement phases, exists.

\section{Monopoles on the lattice and in the continuum}

Let us first  remind the reader the backbone of the theory of the
monopole condensation in the compact photodynamics
\cite{polyakov}. In this case the monopoles are classical
solutions, the same as the original monopoles of Dirac
\cite{dirac}. The radial magnetic field of the monopole is similar
to the electric field of a point-like charge, $|{\bf H}|\sim 1/e^2
r^2$ where $e$ is the electric charge and the factor $1/e^2$
appears because of the Dirac quantization condition. The
corresponding energy is ultraviolet divergent: \beqn
\epsilon_{mon}~\sim \int d^3x \, {\bf H}^2~\sim {1\over e^2 a}\,,
\label{energy} \eeqn where ${\bf H}$ is the magnetic field, $a$ is
the lattice spacing which provides an ultraviolet cut off. Note
that the Dirac string does not contribute to the energy
(\ref{energy}) because of the compactness of the $U(1)$. Otherwise
it would result in a quadratically divergent term (for further
details and references see \cite{main}). Eq. (\ref{energy})
implies that the probability to find a monopole trajectory of
length $L$ is suppressed by the action as $\exp\{-const\cdot
L/(e^2\cdot a)\}$.  This suppression can be overcome, however, for
$e^2\sim 1$ by the entropy factor.  Indeed, on the hypercubic
lattice the number $N$ of trajectories of the length $L$ grows
exponentially, $N\sim\exp(\ln 7\cdot L/a)$ where the constant $\ln
7$ is of pure geometrical origin. Since the self-energy
(\ref{energy}) can be found with all the coefficients fixed the
equating of the entropy and action factors provides a quantitative
means to find the value of $e^2_{crit}$.   A detailed quantitative
analysis along these lines as well as further references can be
found in~\cite{shiba1}.

In case of the gluodynamics, we choose the maximal Abelian gauge
which is defined through maximization of the
functional  $R[U] = \sum_l Tr[\sigma_3 U_l^+ \sigma_3 U_l]$
over all gauge transformations $U_l^\Omega = \Omega^+ U_l \Omega$.
Moreover, in the standard parameterization of the link matrix
\beqn
U_l = \pmatrix{ \cos \varphi_l e^{i\theta_l} & \sin
\varphi_l e^{i \chi_l}\cr - \sin \varphi_l e^{- i \chi_l} & \cos \varphi_l
e^{-i\theta_l} \cr}\,,
\label{LatticeU}
\eeqn
$\theta,\chi\in[-\pi,+\pi)$, $\varphi\in[0,\pi)$,
the functional $R$ can be rewritten as:
$R[U] = \sum_l \cos 2 \varphi_l$.

Thus, the maximization of $R$ corresponds to the
maximization of the absolute values  of the diagonal elements of the link
matrix~\eq{LatticeU}. Since the $SU(2)$ plaquette action is $\beta
\, \frac 12\tr U_P$, at large values of $\beta$ the link matrices
are close to unit matrix up to gauge transformations. Thus, at large values
of $\beta$ in the maximal Abelian gauge $\cos\varphi_l$ are
close to unity, the angles $\varphi_l$ are small and the
$SU(2)$ plaquette action has the form:
\beqn \label{SSU2}
\beta S = \beta \Bigl[ \cos\theta_P \prod\limits^4_{i=1}
\cos \varphi_i + O(\sin\varphi_l)\Bigr].
\eeqn

The projected action $S^{Abel}$ is defined by putting $\varphi_l=0$.
In the form (\ref{SSU2}) the projected action closely resembles the
action of the compact electrodynamics\footnote{Here we neglect
complications due to Faddeev-Popov determinant.}:
\beqn
\label{ScQED}
\beta_{U(1)} S_{cQED} = \beta_{U(1)} \cos \theta_P \, .
\eeqn
However, if we would try to transfer the picture with Abelian monopoles
directly onto the non-Abelian case, the conclusion would be that there
is no monopole condensation in gluodynamics. Indeed,
because of the asymptotic freedom $g^2(a)\to 0$ if $a\to 0$.  Thus,
one substantiates the dual superconductor model of the confinement
with dynamical considerations like the following. Let us start
increasing the lattice spacing a la Wilson. Then the corresponding
effective coupling $g^2$ grows according to the renormgroup
equations. The same coupling governs any of the $U(1)$ subgroups and
once $g^2$ reaches the value where the $U(1)$ monopoles condense (see
above) the condensation occurs in the non-Abelian theory as
well.  In this way one readily understands that there exists
one monopole per volume of order $(\Lambda_{QCD})^{-3}$ so
that the monopoles survive in the continuum limit. However,
since the running of the coupling is a pure quantum effect there
is no much hope to explicitly match a quasi-classical,
Abelian-like field configuration at large distances with perturbative-vacuum
fluctuations at short distances.

Instead, we can think in terms of a phenomenological expansion of the action
density inside the monopoles.
Since the action density is measured in lattice units it
is convenient to consider an expansion of the form:
\beqn
\label{expansion}
& & \big(|F_{\mu\nu}^a|^2_{\mathrm{monopole~center}}-
|F_{\mu\nu}^a|^2_{\mathrm{average}}\big) ~=\nonumber \\
& & = {1\over g^2(a)a^4}\big(
\sum_{k=0}^{\infty} c_k \, g^{2k}(a^2)+\sum_{n=1}^{\infty} b_na^n\big)
\eeqn
Then the theoretical expectation is that all the ultraviolet divergent
pieces vanish, $c_k = 0$. Indeed, otherwise we would have point-like
objects beyond the ordinary gluons, in direct contradiction with the
asymptotic freedom.  Thus, only terms of order $\exp(-const/g^2(a))$ or
powers of $a$ are allowed in the r.h.s. of Eq. (\ref{expansion}).
Moreover, we expect that the series actually starts with the $a^4$
term. Indeed, the monopole field is of order $(F^a_{\mu\nu})_{mon}\sim
\Lambda_{QCD}^2$ as discussed above.  The perturbative fields, on the
other hand, are of order $a^{-2}$.  However, there is no reason to
expect any interference between the perturbative and monopole
contributions, at least upon the averaging.  Thus the excess of the
action near the monopoles is to vanish proportional to $a^4$ if measured
in the lattice units.

The prediction of the $a^4$ behavior holds in the academic limit $a\to 0$.
It is a different matter of course how close to this limit the existing
lattices are. In the next section we will present first indications that
the excess, as measured in the lattice units, decreases with the
decreasing lattice size. However, it is too early to claim that the excess
is vanishing fast at $a\to 0$. In this sense the measurements presented
in this paper can be considered as a first step in studying the monopole
anatomy.

\section{Numerical results}

We have performed measurements of the full non-Abelian action,
$S_{mon}^{SU(2)}$, on the plaquettes closest to the monopole
trajectory. The simulations have been done using Wilson action on
lattices $12^4$ for $\beta= 2.27,\, 2.3,\, 2.33, \, 2.35, \, 2.38, \,
2.4$, $16^4$ for $\beta=2.45$, $20^4$ for $\beta=2.5, 2.55$ and $28^4$
for $\beta=2.6$. We thus kept our physical volume $ \gtrsim 1.5 fm$. We
made $20$ measurements on $12^4$ and $20^4$ lattices, $15$ measurements
on $16^4$ lattice and $20$ measurements on $28^4$ lattice.  The full
non-Abelian action, $S_{mon}^{SU(2)}$, on the plaquettes closest to the
monopole trajectory have been measured.  The error analysis has been
carried out with bootstrap and jackknife methods.  Both methods gave
consistent estimates for statistical errors.

To fix MA gauge the simulated annealing (SA) algorithm was
employed. It is known that this algorithm is vital for reducing
the uncertainty due to Gribov copy effects in the gauge
non-invariant observables computed in MA gauge \cite{bbms}. Our SA
algorithm implementation is essentially the same as described in
\cite{bbms} with the exception that we increased the total number
of SA sweeps up to 2000. To further reduce bias due to Gribov
copies we made gauge fixing for 5 randomly generated gauge copies
for every Monte Carlo configuration. Only the copy with the
maximal value of the gauge fixing functional $R[U]$ has been used
to compute our observables. To estimate the residual effect of the
Gribov copies we compared results obtained with different numbers
of gauge copies $N_{cop}$ in the range from 1 to 5. We have found
systematic albeit weak dependence of our observables on the number
of gauge copies. The difference between results obtained with
$N_{cop}=1$ and $N_{cop}=5$ was less than our statistical errors.
At the same time the difference from results obtained with the
iterative gauge fixing algorithm was an order of magnitude larger
than statistical errors.

\begin{figure}
\epsfxsize=7.5cm
\centerline{\epsfbox{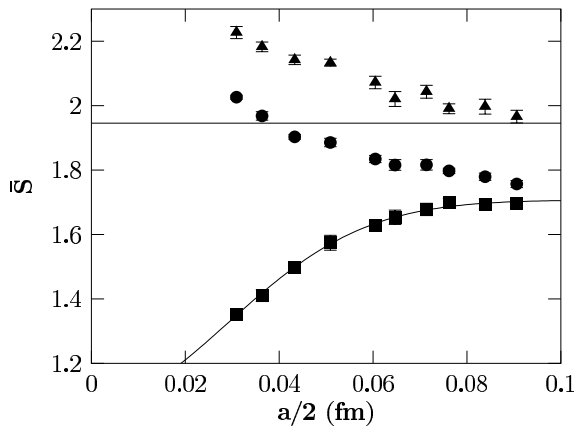}}

\caption{The dependence of excess of the non--Abelian action, $\bar{S}$,
on the distance to the monopole, $a/2$, for all monopoles (circles),
monopoles from IR clusters (boxes) and monopoles from UV clusters
(triangles). The dashed line is $\ln 7$. The error bars are within the
symbols for most of points.}

\label{nonabeact}
\end{figure}

While measuring $S_{mon}$ on the plaquettes closest to the
monopole trajectory, we discriminate between the monopoles
belonging to the IR and UV clusters.  Our lattices are of the
physical size $ \gtrsim 1.5fm$ , i.e. large enough for most of
observables. On the other hand it has been found out in
\cite{percolation} that essentially larger volume is necessary to
have only one large IR cluster. Some of our lattices are not large
enough and we find on them not one but a few large clusters. We
believe that combining all these clusters one gets the set of
infrared monopoles. This conjecture has been confirmed recently
\cite{Bornyakov:2001ux}. In the present work we considered only
monopoles from the largest cluster as IR monopoles. This
introduces some systematic uncertainty into results for UV
clusters, namely the corresponding non-Abelian action is
underestimated. At the same time this uncertainty does not
influence our main conclusions.

In Fig.~\ref{nonabeact} we show the dependence of
$\bar{S} = 6\beta\,\langle S_{mon}\, - \,S \rangle$ on the half of the
lattice spacing\footnote{To define the lattice distance in Fermi, we
find the correspondence between the bare charge and lattice spacing by
fixing the value of the string tension $\sigma = 440\,\, \mbox{MeV}$ and
using the numerical data for the string tension in lattice units,
$\sigma\cdot a^2$, see~\cite{corr}.}, $a/2$.  The factor $6\beta$ is
introduced here to make convenient the comparison of the action and the
entropy factors. The explanation of the scale of the horizontal axis,
i.e.  $a/2$, is the following. Since $\langle S_{mon} \rangle$ is
measured on the plaquettes which are faces of the cube dual to the
monopole current, this corresponds to measuring the average field
strength at the distance $a/2$ from the monopole center. Note that the
excess of the action is dominated by the closest plaquettes.
In this figure
we compare the action on the plaquettes nearest to the monopole
center with the $\ln 7$. As is mentioned in Sect. 2, the $\ln 7$ is a
geometrical constant determining the monopole entropy. The action
in the lattice units for the percolating monopoles should not
exceed $\ln{7}$, see, e.g., \cite{shiba2,shiba1} and references
therein.

\begin{figure}
\epsfxsize=7.5cm
\centerline{\epsfbox{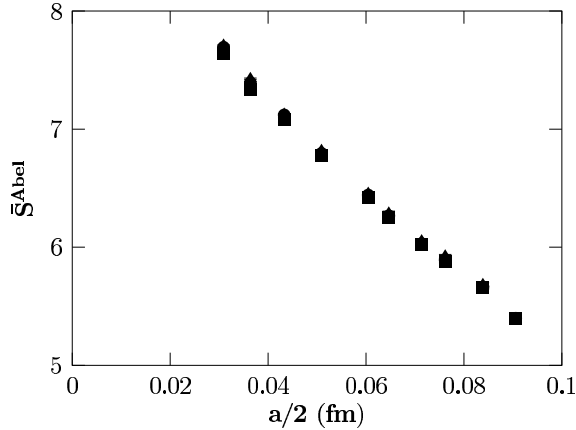}}
\caption{The same as in Fig.~\ref{nonabeact} for $\bar{S}^{Abel}$.}
\label{abeact}
\end{figure}

Our main observation is that $\bar{S}$ for the monopole belonging
to the IR cluster decreases when we approach the monopole center.
Moreover, it is below $\ln 7$ for all data in agreement with
percolation condition discussed above.  On the other hand, the
action for the UV monopoles is increasing and exceeds $\ln 7$ in
agreement with the fact that these clusters are not percolating.
Thus for the first time we demonstrate by direct computation of
the action density that the percolation condition works in $SU(2)$
gluodynamics: percolating monopoles carry action density (in
lattice units) less than $\ln 7$, while for non-percolating
monopoles the action density is above this value. Note, that the
excess of the action near all monopoles behaves similarly to the
leading term in the monopole action~\cite{SuzukiPPP}. This term is
proportional to the length of the monopole trajectory.

The action density distribution on Fig.~\ref{nonabeact} has the same
physical meaning as the action profile of the 't~Hooft--Polyakov
monopole classical solution~\cite{thooftpolyakov}.  However the
monopoles studied in the present Letter are of a purely quantum origin.

The results of the calculation of the Abelian action near the
monopole, $\bar{S}^{Abel} = 6\beta\, \langle S^{Abel}_{mon}\, -
\,S^{Abel}\rangle$, are presented in Fig.~\ref{abeact}. Unlike the
full non-Abelian action, the Abelian action associated with the
monopoles, $\bar{S}^{Abel}$, for monopoles belonging to IR and UV
clusters is approximately the same. Moreover, it increases when
one approaches the center of monopole. There is no known
explanation of this effect. A comparison of Figs.~\ref{nonabeact}
and \ref{abeact} shows that the role of the off--diagonal degrees
of freedom seem to compensate the divergent contribution into the
monopole energy from the Abelian part of the gluon fields. Thus
the monopoles in the maximal Abelian projection look like 't Hooft
-- Polyakov monopoles which are not singular at the origin. Note
that the monopoles in compact QED are singular Dirac monopoles.

Now we discuss the size of the IR monopole. It occurs that the fit
of the data in Fig.~\ref{nonabeact} by the function $C_0 + C_1
\exp\{- R^2 \slash {(R^{mon})}^2\},\, R=a/2$, can be performed with
high quality, $\chi^2/N_{DOF} = 0.26$. The values of fit
parameters are: $C_0=1.706(5)$, $C_1 = -0.63(2)$, $R^{mon} =
0.041(1) \, fm$.  This fit is shown in Fig.~\ref{nonabeact}
by solid line. Thus our estimation for the monopole radius is
$R^{mon}\approx 0.04\, fm$.  Of course the definition of the
monopole radius is not unique, but we believe that all reasonable
definitions give the monopole radius of the same order. Note that
in Ref.~\cite{shiba2} it was found that the monopole condensation
starts for monopoles approximately of the same physical size as
$R^{mon}$ determined in the present paper. Up to now we did not
study the scaling behaviour of the monopole radius.  Such
calculations are possible if we surround monopoles by cubes of
various size and we measure the action on the faces of these
cubes. This study is now in progress.

Our data give support to the following picture.  At the large enough
distances from the monopoles gauge field is Abelian-like and
approximation eq.~\eq{SSU2} works well. At short distances the
non-Abelian nature of the monopoles is manifest and while the Abelian
part of the action grows up the total action decreases.

To summarize, we have shown that the phenomenon of the monopole
condensation in the lattice gluodynamics is due to a special anatomy of
the monopoles belonging to the IR cluster. On the other hand, in the
limit $a\to 0$ one would expect much faster vanishing of the excess of
the action than it was observed so far.  Since the theoretical
prediction on the vanishing of the excess of the action at small $a$
seems very reliable (see the preceding section) the results obtained are
to be rather interpreted in terms of various scales of the
non-perturbative physics. Indeed, the average distance between nearest
monopoles in the IR cluster is about $0.5\,\, fm$, as can be extracted from
the data in Refs. \cite{percolation,Bornyakov:2001ux}. Now we observe
for the first time that the excess of the action goes down when we
approach the monopole center. The corresponding radius turns out to be
small numerically\footnote{Note that this implies that monopoles form a
gas rather than a liquid since the monopole cores are most probably not
interacting due to the separation of scales.}, $R^{mon}\approx
0.04~fm$. Moreover, even a smaller scale might emerge in future. Indeed,
in the limit $a\to 0$ we should have the $a^4$ behavior for the excess
of the action which is not yet in sight at present.  Thus, there appears
a hierarchy of scales all of which are formally of the same order, $\sim
\Lambda_{QCD}$. Note that existence of such hierarchies has already been
conjectured on various grounds.  First, a great variety of scales is
manifested through QCD sum rules \cite{sumrules}.  Further evidence has
been accumulated via various lattice measurements, see
\cite{scales} and references therein. In particular, very
recently the relevance of the scale of order $2~\mathrm{GeV}$ was
revealed through the measurements of the $\langle A^2 \rangle$ vacuum
condensate.  This scale would roughly correspond to the monopole radius
$R^{mon}\approx 0.04 fm$ which we are observing.

\section*{Acknowledgments}

M.I.P. and A.I.V. acknowledge the kind hospitality of the staff of the
Institute for Theoretical Physics of Kanazawa University, where the
work was initiated. Work of V.G.B., M.N.C., F.V.G., M.I.P. and
A.I.V. was partially supported by grants RFBR 02-02-17308, RFBR
01-02-17456, INTAS 00-00111, JSPS Grant in Aid for Scientific Research
(B) (Grant No. 10440073 and No. 11695029) and CRDF award RP1-2364-MO-02.

\end{document}